\documentclass[12pt]{article}
\pdfoutput=1
\usepackage{graphicx,epstopdf,amssymb,amsfonts,amsmath,amsthm,array,
mathrsfs,amscd}
\usepackage{todonotes} 
\newcommand{\pbs}[1]{\let\temp=\\#1\let\\=\temp}
\numberwithin{equation}{section}
\def\be{\begin{equation}}\def\ee{\end{equation}}
\def\cvp{\raise 2pt\hbox{,}}

 \def\d{{\rm d}}

\def\d{\partial}

\def\d{{\rm d}}

\def\g{\gamma}
\def\G{\Gamma}
\def\dd{\delta}

\def\m{\mu}

\def\l{\lambda}

\def\s{\sigma}
\def\f{\phi}

\def\D{\Delta}

\def\vf{\varphi}
\def\L{\Lambda}

\def\wt{\widetilde}

\def\l{\lambda}

\def\del{\partial}

\def\ba{\begin{eqnarray}}
\def\ea{\end{eqnarray}}

\theoremstyle{plain}

\theoremstyle{definition}
\theoremstyle{remark}

\def\imath#1#2#3{{\it Invent math }{\bf #1} (#2) #3}

\begin{document}
%
%


{\pagestyle{empty}
\parskip 0in
\

\vfill
\begin{center}

{\Large \bf 2D quantum gravity on compact Riemann surfaces}

\medskip

{\Large \bf with non-conformal matter}

\vspace{0.4in}

Adel B{\scshape ilal}$^{*}$  
and L\ae titia L{\scshape educ}$^{\dagger}$
\\

\medskip
$^{*}$\it {Laboratoire de Physique Th\'eorique de l'\'Ecole Normale Sup\'erieure\\
PSL Research University, CNRS, Sorbonne Universit\'es, UPMC Paris 6\\
24 rue Lhomond, F-75231 Paris Cedex 05, France}

\medskip

$^{\dagger}$\it {Institut f\"ur Theoretische Physik, Universit\"at zu K\"oln\\
Z\"ulpicher Stra\ss e 77, 50937 K\"oln, Germany}

\smallskip
\end{center}
\vfill\noindent
We study  the gravitational action induced by coupling two-dimensional non-conformal, massive matter  to gravity on a compact Riemann surface. We express this gravitational action in terms of finite and well-defined quantities for any  value of the mass. A small-mass expansion gives back the Liouville action in the massless limit, the Mabuchi and Aubin-Yau actions to first order, as well as an infinite series of higher-order contributions written in terms of purely geometric quantities.
\vfill
\medskip
%
\begin{flushleft}
\end{flushleft}
%
\hrule

{\parskip -0.3mm
\small{\tableofcontents}}
}
\newpage
\setcounter{page}{1}


%
%
\section{Introduction and generalities}

\subsection{Introduction and motivation}

Ever since the seminal paper by Polyakov \cite{Liouville1} it has been known that conformal matter coupled to 2D gravity gives rise to an ``effective" gravitational action that is the Liouville action
\be\label{Liouvaction}
S_L[g_0,g] \equiv S_L[g_0,\s]=\int\d^2 x \sqrt{g_0} \, \big( \s \D_0\s + R_0 \s\big) 
, \quad g=e^{2\s} g_0\ .
\ee
More precisely, the Liouville action captures the dependence on the metric of the partition function $Z_{\rm mat}^{(c)}$ of conformal matter of central charge $c$. If we consider two metrics $g_0$ and $g=e^{2\s}g_0$ then
\be\label{Sgravconfmat}
-\ln \frac{Z_{\rm mat}^{(c)}[g]}{Z_{\rm mat}^{(c)}[g_0]}=\frac{c}{24\pi} S_L[g_0,g] \ .
\ee
For more general ``matter" (plus ghost) partition functions one defines a general gravitational action as
\be\label{Sgravgen}
-\ln \frac{Z_{\rm mat}[g]}{Z_{\rm mat}[g_0]}= S_{\rm grav}[g_0,g] \ .
\ee
Being defined as the logarithm of a ratio of partition functions computed with two different metrics, any gravitational action satisfies a cocycle identity 
\be\label{cocycle}
S_{\rm grav}[g_1,g_2] + S_{\rm grav}[g_2,g_3]=S_{\rm grav}[g_1,g_3] \ .
\ee
The simplest example of a gravitational action satisfying this cocycle identity is the ``cosmological constant action"
\be\label{cosmolconst}
S_c[g_0,g]=\m_0  \int \d^2 x (\sqrt{g} -\sqrt{g_0})=\m_0 (A-A_0) \ .
\ee
As we will recall below, this action must be present as a counterterm to renormalize the divergences that are present in \eqref{Sgravconfmat} in addition to $S_L$.
Gravitational actions other than the Liouville or cosmological constant actions can be constructed and have been studied mainly in the mathematical literature, like the Mabuchi and Aubin-Yau actions \cite{Mabuchi,AubinYau}. These latter functionals  crucially involve not only the conformal factor $\s$ but also directly the K\"ahler potential $\f$ and do admit generalizations to higher-dimensional  K\"ahler manifolds. In the mathematical literature they  appear in relation with the characterization of constant scalar curvature metrics \cite{AubinYau}.
Their r\^oles as two-dimensional gravitational actions in the sense of \eqref{Sgravgen} have been discussed in some detail in \cite{FKZ}. In particuler, ref.~\cite{FKZ} has studied the metric dependence of the partition function of non-conformal matter like a massive scalar field and shown that  a gravitational action defined by \eqref{Sgravgen} contains these Mabuchi and Aubin-Yau actions  as first-order corrections (first order in $m^2 A$ where $m$ is the mass and $A$ the area of the Riemann surface) to the Liouville action.
The partition function of quantum gravity at fixed area, with a gravitational action being a combination of the Liouville and Mabuchi actions, has been studied at one loop in \cite{BFK} and at  two and three loops in \cite{BL}.

While the results of \cite{FKZ}  concerned the first order in an expansion in $m^2 A$, in the present note we will  derive a few results that are valid exactly at finite $m$. Ideally one would like to study some general matter action where non-conformal terms $\sim c_i {\cal O}^i$ have been added to some conformal theory and obtain exact results in these couplings $c_i$. We are going to be much less ambitious and simply study a single massive scalar field with action
\be\label{scalaraction}
S_{\rm mat}[g,X]=\frac{1}{2}\int\d^2 x \sqrt{g} \left[ g^{ab}\del_a X \del_bX + m^2 X^2\right]
=\frac{1}{2}\int\d^2 x \sqrt{g}\, X (\D_g + m^2)X  \ ,
\ee
and try to establish some exact results valid for finite $m$.
Here $\D_g$ is the Laplace operator for the metric $g$, defined with a minus sign, so that its eigenvalues are non-negative:
\be\label{Laplace}
\D_g=-\frac{1}{\sqrt{g}}\del_a(g^{ab}\sqrt{g} \del_b) \ .
\ee

In the remainder of this section, we discuss some basic differences between the massive and massless cases and define the matter partition functions in both cases. Then we briefly recall the Mabuchi and Aubin-Yau actions and their variations. In section 2, we summarize some  technical tools involving heat kernels, zeta functions and  their perturbation theory. Section 3 then gives the computation of the gravitational action for the massive scalar field, providing explicit formulae for its dependence on the metric in terms of mani\-festly finite and well-defined quantities. 
Maybe not too surprisingly, some of our massive formulae  will look somewhat similar to those that can be found in \cite{FKZ} for the massless case. However, let us insist that our results are exact in $m$ and valid for any finite mass. Nevertheless, we will write them  in a way that immediately allows for a small mass expansion, thus recovering the Liouville action in the zero-mass limit and the Mabuchi and Aubin-Yau actions as the first-order corrections. The higher-order corrections are similarly expressed in terms of purely geometric objects, but do not seem to have any known counterparts in the mathematical literature.

\subsection{Massive versus massless matter\label{massivemassless}}

One should keep in mind that adding the mass term is {\it not} just a perturbation by some operator  that has a non-zero conformal weight. This is due to the (would-be) zero-mode of the scalar field that is absent from the action for zero mass but obviously plays an important r\^ole for non-zero mass. In particular, this means that the relevant quantities of the massive theory are not simply given by those of the massless theory plus order $m^2$ corrections. This is most clearly examplified by the Green's function $G(x,y)$ of the operator $\D_g+m^2$. The latter is hermitian and has a complete set of eigenfunctions $\vf_n$ with eigenvalues $\l_n\ge m^2$~:
\be\label{eigenvv}
(\D_g+m^2) \vf_n(x) =\l_n \vf_n(x) \quad , \quad n=0,1,2,\ldots \ .
\ee
Clearly, the eigenfunctions $\vf_n$ do not depend on $m$ (i.e. $\vf_n=\vf_n^{(0)}$), while $\l_n=\l_n^{(0)}+m^2$.   The eigenfunctions which may be chosen to be real, are orthonormalized as
\be\label{orthonormal}
\int\d^2 x \sqrt{g} \,\vf_n(x)\vf_k(x)=\dd_{nk} \ .
\ee
As is clear from  \eqref{Laplace}, $\D_g$ always has a zero mode and, hence,
\be\label{zeromode}
\l_0= m^2 \quad , \quad \vf_0=\frac{1}{\sqrt{A}} \ .
\ee
We always refer to $\vf_0$ as the zero-mode, even in the massive case.
The Green's function for $m\ne 0$ is given by
\be\label{Green}
G(x,y)=\sum_{n\ge 0} \frac{\vf_n(x)\vf_n(y)}{\l_n}
\quad , \quad
(\D_g+m^2) G(x,y) = \frac{1}{\sqrt{g}} \dd(x-y) \ .
\ee
In general, if $B$ is any quantity defined for $m\ne 0$, we will denote by $B^{(0)}$ the corresponding quantity for $m=0$. But, for $m=0$, the Green's function is not simply $G^{(0)}$ since  $\l_0^{(0)}=0$ and, obviously, the zero-mode must be excluded from the sum. Then
\be\label{Greenm=0}
\wt G^{(0)}(x,y)=\sum_{n> 0} \frac{\vf_n(x)\vf_n(y)}{\l_n^{(0)}}
\quad , \quad
\D_g \wt G^{(0)}(x,y) = \frac{1}{\sqrt{g}} \dd(x-y) -\frac{1}{A}\ .
\ee
The subtraction of $\frac{1}{A}$ on the r.h.s.~ensures that, when integrated $\int\d^2 x\sqrt{g}\ldots$, one correctly gets zero. We will consistently put a tilde over the various quantities we will encounter if the zero-mode is excluded from the sum.\footnote{
Except for determinants missing the zero-mode, where we will write $\det'$ following the usual notation.} In particular, using \eqref{zeromode}, we can write for the massive Green's function 
\be\label{GminusGtilde}
G(x,y)=\frac{1}{m^2 A} + \wt G(x,y)
\quad , \quad
(\D_g+m^2) \wt G(x,y) = \frac{1}{\sqrt{g}} \dd(x-y) -\frac{1}{A} \ ,
\ee
where $\wt G$ has a smooth limit as $m\to 0$. Moreover,
the smallest eigenvalue contributing in $\wt G$ is $\l_1=\l_1^{(0)}+m^2$ with $\l_1^{(0)}>0$ being of order $\frac{1}{A}$. Thus, if $A m^2\ll 1$, one can expand $\frac{1}{\l_n}=\sum_{r=0}^\infty (-)^r \frac{m^{2r}}{(\l_n^{(0)})^{r+1}}$, resulting in
\be\label{massiveGexp}
G(x,y)=\frac{1}{m^2 A} + \wt G^{(0)}(x,y) + \sum_{r=1}^\infty (-m^2)^r \wt G^{(0)}_{r+1}(x,y) \ ,
\ee
where
\be\label{Gn}
G_{r}(x,y)=\sum_{n\ge 0}\frac{\vf_n(x)\vf_n(y)}{\l_n^r} \quad , \quad
\wt G_{r}^{(0)}(x,y)=\sum_{n> 0}\frac{\vf_n(x)\vf_n(y)}{(\l_n^{(0)})^r} \quad .
\ee
Clearly, the massive Green's function does not equal the massless one plus order-$m^2$ corrections: their is a crucial $\frac{1}{m^2 A}$ term in \eqref{GminusGtilde} and \eqref{massiveGexp}.

The matter partition function is defined  with respect to the decomposition  $X=\sum_{n\ge 0} c_n \vf_n$  as
\be\label{Zmat}
Z_{\rm mat}[g]=\int{\cal D}_g X e^{-S_{\rm mat}[g,X]} 
= \int \prod_{n=0}^\infty \frac{\d c_n}{\sqrt{2\pi}} e^{-\frac{1}{2} \sum_{n\ge 0} \l_n c_n^2}
=\big(\det (\D_g+m^2) \big)^{-1/2} .
\ee
In the massless case, since $\l_0^{(0)}=0$, the integration over $c_0$ would be divergent  and instead one replaces it by a factor $\sqrt{A}$. Thus
\be\label{Zmat0}
Z^{(0)}_{\rm mat}[g]=\int{\cal D}_g^{(0)} X e^{-S_{\rm mat}^{(0)}[g,X]} 
= \sqrt{A} \int\prod_{n=1}^\infty \frac{\d c_n}{\sqrt{2\pi}} e^{-\frac{1}{2}\sum_{n> 0} \l_n^{(0)} c_n^2}=\Big(\frac{\det' \D_g}{A} \Big)^{-1/2} .
\ee
Of course, the determinants $\det$ and $\det'$ are ill-defined and need to be regularized. We will use the very convenient regularization-renormalization in terms of the spectral $\zeta$-functions:
\be\label{zeta}
\zeta(s)=\sum_{n=0}^\infty \l_n^{-s}
\quad , \quad 
\wt\zeta(s)=\sum_{n=1}^\infty \l_n^{-s} \ ,
\ee and similarly for $\wt\zeta^{(0)}(s)$. By Weil's law (see e.g.~\cite{BF}), the asymptotic behaviour of the eigenvalues for large $n$ is $\l_n \sim \frac{n}{A}$ and, hence the spectral  $\zeta$-functions are defined by converging sums for ${\rm Re}\, s>1$, and by analytic continuations for all other values. In particular, they are well-defined meromorphic functions for all $s$ with a single pole at $s=1$ with residue $\frac{1}{4\pi}$ (see e.g. \cite{BF}). A straightforward formal manipulation shows that $\zeta'(0)\equiv \frac{\d}{\d s}\zeta(s)\vert_{s=0}$ provides a formal definition of $-\sum_{n\ge 0} \ln \l_n$, i.e. of $-\ln \det(\D_g+m^2)$:
\be\label{Zmatzeta}
Z_{\rm mat}[g]=\exp\Big({\frac{1}{2} \zeta'(0)}\Big) \quad, \quad
Z^{(0)}_{\rm mat}[g]=A^{1/2}\ \exp\Big({\frac{1}{2} (\wt\zeta^{(0)})'(0)}\Big) \ .
\ee

There is a slight subtlety one should take into account, see e.g. \cite{BF}. While the field $X$ is dimensionless, the $\vf_n$ scale as $A^{-1/2}\sim \m$ where $\m$ is some arbitrary mass scale (even if $m=0$), and the $c_n$ as $\m^{-1}$. It follows that one should write ${\cal D}_g X=\prod_n \frac{\m\d c_n}{2\pi}$. This results in $Z_{\rm mat}=\big(\prod_n \frac{\l_n}{\m^2}\big)^{-1/2}$, so that every $\zeta'(0)$ is changed into
\be\label{zetamu}
\zeta'(0) \to \zeta'(0)+ \zeta(0)\, \ln\m^2 \ .
\ee

The regularization-renormalization of determinants in terms of the $\zeta$-function may appear as rather ad hoc, but it can be rigorously justified by introducing the spectral regularization \cite{BF}. The regularized logarithm of the determinant then equals $\zeta'(0)+\zeta(0)\ln \m^2$ plus a diverging piece $\sim A\L^2 (\ln\frac{\L^2}{\m^2} + {\rm const})$, where $\L$ is some cutoff. This diverging piece just contributes to the cosmological constant action \eqref{cosmolconst}, and this is why the latter must be present as a counterterm, to cancel this divergence.

Thus, finally
\ba\label{Sgravgen2}
\hskip-1.cm S_{\rm grav}[g_0,g]&=&-\frac{1}{2} \left(\zeta_g'(0)+\zeta_g(0) \ln\m^2\right) +\frac{1}{2} \left(\zeta_{g_0}'(0)  +\zeta_{g_0}(0)\ln\m^2\right)\, 
\ , 
\nonumber\\
\hskip-1.cm S_{\rm grav}^{(0)}[g_0,g]
&=&-\frac{1}{2}\ln\frac{A}{A_0}-\frac{1}{2} \left( (\wt\zeta_g^{(0)})'(0)+\wt\zeta_g^{(0)}(0) \ln\m^2\right)
+\frac{1}{2}\left( (\wt\zeta_{g_0}^{(0)})'(0) + \wt\zeta_{g_0}^{(0)}(0)\ln\m^2
\right) 
\, ,
\ea
where the first line refers to the massive case and the second line to the massless one.

\subsection{Mabuchi and Aubin-Yau actions}

Let us briefly recall the basic properties of the known gravitational actions.
While the Liouville action \eqref{Liouvaction} can be written in terms of $g_0$ and the conformal factor $\s$, the Mabuchi and Aubin-Yau actions are formulated using also the K\"ahler potential $\f$. They are related by
\be\label{gg0sig}
g = e^{2\sigma}g_{0} \quad , \quad
e^{2\s}= \frac{A}{A_0}\left(1-\frac{1}{2} A_0 \D_0\f\right)\ ,
\ee
where $\D_0$ denotes the Laplacian for the metric $g_0$ with area $A_0=\int\d^2 x \sqrt{g_0}$. The Mabuchi action on a Riemann surface of genus $h$ can then be written as \cite{FKZ}
\be\label{Mab1}
S_{\rm M}[g_0,g]
=\int\d^2 x \sqrt{g_0} \left[ 2\pi(h-1)\f\D_0\f + \Bigl(\frac{8\pi(1-h)}{A_0}-R_0\Bigr) \f +\frac{4}{A} \s e^{2\s} \right] \  ,
\ee
while the Aubin-Yau action takes the form
\be\label{AubY}
S_{\rm AY}[g_0,g]=-\int\d^2 x \sqrt{g_0} \left[ \frac{1}{4} \f\D_0\f -\frac{\f}{A_0}\right] \ .
\ee
As already mentioned, they both satisfy a cocycle identity analogous to \eqref{cocycle} and were shown \cite{FKZ} to appear in $S_{\rm grav}$ in the term of first order in an expansion  in $m^2 A$. Note that $S_{\rm M}=8\pi(1-h) S_{\rm AY}+\int\d^2 x \sqrt{g_0}\left( \frac{4}{A} \s e^{2\s} -R_0\f\right)$. Eq.~\eqref{gg0sig} relates the variations $\dd\s$ and $\dd\f$ as 
\be\label{deltasigmadeltaphi}
\dd\s=\frac{\dd A}{2A}-\frac{A}{4} \D\dd\f
\quad \text{and}\quad
\dd\left(\frac{e^{2\s}}{A}\right)=-\frac{1}{2}\D_0\dd\f \ .
\ee
It is then straightforward to show that the variations of the Liouville, Mabuchi and Aubin-Yau actions are given by
\ba\label{SMSAYvar}
\dd S_L[g_0,g]&=&4\pi(1-h)\frac{\dd A}{A} -\frac{A}{4} \int\d^2 x \sqrt{g}\, \D R\, \dd\f \ ,
\nonumber\\
\dd S_{\rm M}[g_0,g]&=&2\frac{\dd A}{A}-\int\d^2 x \sqrt{g}\, \left( R-\frac{8\pi(1-h)}{A}\right) \dd\f \ ,
\nonumber\\
\dd S_{\rm AY}[g_0,g]&=&\frac{1}{A}\int\d^2 x \sqrt{g}\, \dd\f \ .
\ea
Thus the Liouville and Mabuchi actions obviously admit the constant scalar curvature metrics as saddle-points at fixed area. Although not obvious from the previous equation, the variation of the Aubin-Yau action when restricted to the space of Bergmann metrics is similarly related to metrics of constant scalar curvature \cite{AubinYau}.

\section{Some technical tools}

\subsection{The heat kernel}

The heat kernel and integrated heat kernel for the operator $\D_g+m^2$ are defined in terms of the eigenvalues and eigenfunctions \eqref{eigenvv} as
\be\label{heatkernel}
K(t,x,y)=\sum_{n\ge0} e^{-\l_n t} \,\vf_n(x) \vf_n(y) 
\quad , \quad
K(t)=\int\d^2 x \sqrt{g}\, K(t,x,x)=\sum_{n\ge 0} e^{-\l_n t} \ .
\ee
The corresponding $\wt K$, $K^{(0)}$ and $\wt K^{(0)}$ are defined similarly.
The heat kernel $K$ is the solution of
\be\label{heatdiffeq}
\left(\frac{\d}{\d t} + \D_g+m^2\right) K(t,x,y) = 0 
\quad , \quad
K(t,x,y)\sim\frac{1}{\sqrt{g}} \dd(x-y)\ \text{as\ } t\to 0 \ .
\ee
Note that it immediately follows from either \eqref{heatkernel} or \eqref{heatdiffeq} that the massless and massive heat kernels are simply related by
\be\label{KmKo}
K(t,x,y)=e^{-m^2 t} K^{(0)}(t,x,y) \ .
\ee
As is also clear from  \eqref{heatkernel} (and Weil's law $\l_n\sim\frac{n}{A}$), for $t>0$, $K(t,x,y)$ is given by a converging sum and is finite, even as $x\to y$. For $t\to 0$ one recovers various divergences, in particular 
\be\label{KG}
\int_0^\infty \d t \, K(t,x,y) = G(x,y)
\ee
exhibits the short distance singularity of the Green's function which is well-known to be\break logarithmic.

The behaviour of $K$ for small $t$ is related to the asymptotics of the eigenvalues $\l_n$ for large $n$, which in turn is related to the short-distance properties of the Riemann surface. It is thus not surprising that the small-$t$ asymptotics is given in terms of local expressions of the curvature and its derivatives. Indeed, one has the well-known small $t$-expansion :
\be\label{Kasymp}
K(t,x,y) \sim \frac{1}{4\pi t} e^{-(\ell^2/4t) - m^2 t} \Big[ a_0(x,y)+ t \,a_1(x,y) + t^2\, a_2(x,y) +\ldots \Big] 
\ee
where $\ell^2\equiv\ell^2_g(x,y)$ is the geodesic distance squared between $x$ and $y$ as measured in the metric $g$. For small $t$, the exponential forces $\ell^2$ to be small (of order $\sqrt{t}$) and one can use normal coordinates around $y$. This allows one to obtain quite easily explicit expressions for the $a_r(x,y)$ in terms of the curvature tensor and its derivatives. They can be found e.g. in \cite{BF}. Here, we will only need them  at coinciding points $y=x$, where
\be\label{Kxx}
K(t,x,x)\sim  \frac{1}{4\pi t} \Big[1 + \Big(\frac{R}{6}-m^2\Big) \, t + \ldots\Big] \ .
\ee
Let us note that in the massless case and if the zero-mode is excluded one has instead
$\wt K^{(0)}(t,x,x)\sim  \frac{1}{4\pi t} \Big[1 + \Big(\frac{R}{6}-\frac{4\pi}{A}\Big) \, t + \ldots\Big]$.

\subsection{Local $\zeta$-functions and Green's function at coinciding points}

Local versions of the $\zeta$-functions are defined as
\be\label{localzeta}
\zeta(s,x,y)=\sum_{n\ge0} \frac{\vf_n(x) \vf_n(y) }{\l_n^s} \ ,
\ee
and similarly for $\wt\zeta(s,x,y)$, etc. Note that $\zeta(1,x,y)=G(x,y)$, while for $s=r=2,3,\ldots$ these local $\zeta$-functions coincide with the $G_r(x,y)$ defined above in \eqref{Gn}. They are related to the heat kernel by
\be\label{zetaheat}
\zeta(s,x,y)=\frac{1}{\G(s)} \int_0^\infty \d t\, t^{s-1} K(t,x,y) \ .
\ee
For $s=0,-1,-2,\ldots$, $\G(s)$ has poles and the value of $\zeta(s,x,y)$ is entirely determined by the singularities of the integral over $t$ that arise from the small-$t$ asymptotics of $K$. As shown above, the latter is given by local quantities on the Riemann surface. In particular,
\be\label{zetaat0}
\zeta(0,x,x)=\frac{R(x)}{24\pi}-\frac{m^2}{4\pi} 
\quad \text{and}\quad
\wt\zeta(0,x,x)=\frac{R(x)}{24\pi}-\frac{m^2}{4\pi} -\frac{1}{A}\ .
\ee
On the other hand, the values for $s=1,2,3,\ldots$ or the derivative at $s=0$ cannot be determined just from the small-$t$ asymptotics and require the knowledge of the full spectrum of $\D_g+m^2$, i.e. contain global information about the Riemann surface.

Clearly, $\zeta(1,x,y)=G(x,y)$ is singular as $x\to y$. For $s\ne 1$,  $\zeta(s,x,y)$  provides a regularization of the propagator. More precisely,  it follows from \eqref{zetaheat} that $\zeta(s,x,x)$ is a meromorphic function with a pole at $s=1$ and that the residue of this pole  is $\frac{a_0(x,x)}{4\pi}=\frac{1}{4\pi}$. Thus  \cite{BF}
\be\label{Gzetazetadef}
G_\zeta(x)=\lim_{s\to 1}\left[ \m^{2(s-1)} \zeta(s,x,x)-\frac{1}{4\pi(s-1)}\right] 
\ee
is well-defined. (Here $\m$ is an arbitrary scale.) This is an  important quantity, called the ``Green's function at coinciding points". One can give an alternative definition of $G_\zeta$ by subtracting the short distance singularity from $G(x,y)$ and taking $x\to y$. More precisely \cite{BF}
\be\label{Gzetadef}
G_\zeta(y)=\lim_{x\to y} \left[ G(x,y) + \frac{1}{4\pi}\left( \ln\frac{\ell^2_g(x,y) \m^2}{4} +2\g\right)\right] \ .
\ee
One can show \cite{BF} that both definitions of $G_\zeta$ are equivalent and define the same quantity. Again, the same relations hold between $\wt G_\zeta(y)$, $\wt G(x,y)$ and $\wt \zeta(s,x,x)$. Note that $G_\zeta(y)$ contains global information about the Riemann surface and cannot be expressed in terms of local quantities only.

\subsection{Perturbation theory}

We want to study how the eigenvalues $\l_n$ and eigenfunctions $\vf_n$ change under an infinitesimal change of the metric. Since $g=e^{2\s}g_0$, the Laplace operator $\D_g$ and hence also $\D_g+m^2$ only depend on the conformal factor $\s$ and on $g_0$:
$\D_g=e^{-2\s}\D_{0}$ and thus under a variation $\dd\s$ of $\s$ one has
\be\label{Lapvar}
\dd \D_g=-2\dd\s \D_g \quad \Rightarrow\quad
\langle \vf_k | \dd \D_g | \vf_n\rangle = -2(\l_n-m^2) \langle \vf_k | \dd \s | \vf_n\rangle \ ,
\ee
where, of course, $\langle \vf_k | \dd \s | \vf_n\rangle=\int\d^2 x \sqrt{g}\, \vf_k \dd\s \vf_n$. 
One can then apply standard quantum mechanical perturbation theory. The only subtlety comes from the normalisation condition \eqref{orthonormal} which also gets modified when varying $\s$ \cite{FKZ,BF}. One finds
\ba\label{deltalambda}
\dd\l_n&=&-2(\l_n-m^2) \langle \vf_n | \dd \s | \vf_n\rangle \ ,
\\
\label{deltaphi}
\dd\vf_n&=&- \langle \vf_n | \dd \s | \vf_n\rangle\, \vf_n - 2 \sum_{k\ne n} \frac{\l_n-m^2}{\l_n-\l_k} \langle \vf_k | \dd \s | \vf_n\rangle \, \vf_k \ .
\ea
Let us insists that this is first-order perturbation theory in $\dd\s$, but it is exact in $m^2$.
Note the trivial fact that, since $\l_0=m^2$ and $\vf_0=\frac{1}{\sqrt{A}}$, one has consistently
\be\label{deltalamzero}
\dd\lambda_0=0 \quad , \quad \dd\vf_0=\dd\big( \frac{1}{\sqrt{A}}\big) \ .
\ee
\section{The gravitational action}

\subsection{Variation of the determinant}

Recall from \eqref{Sgravgen2} that the gravitational action is defined as
$S_{\rm grav}[g_0,g]=-\frac{1}{2} \left(\zeta_g'(0)-\zeta_{g_0}'(0)\right)\break -\frac{1}{2} \left(\zeta_g(0)-\zeta_{g_0}(0)\right)\ln\m^2$. Our goal is to compute $\dd \zeta'(0)\equiv \dd \zeta_g'(0)$ and $\dd \zeta(0)\equiv \dd \zeta_g(0)$ and express them as  ``exact differentials" so that one can integrate them and obtain the finite differences $\zeta_{g_2}'(0)-\zeta_{g_1}'(0)$ and $\zeta_{g_2}(0)-\zeta_{g_1}(0)$.

From \eqref{deltalambda} one immediately gets, to first order in $\dd\s$,
\be\label{deltazeta}
\zeta_{g+\dd g}(s)=\sum_{n\ge 0}\frac{1}{(\l_n+\dd\l_n)^s}
=\zeta_g(s) + 2 s \sum_{n\ge 0} \frac{\l_n-m^2}{\l_n^{s+1}}  \langle \vf_n | \dd \s | \vf_n\rangle \ ,
\ee
As noted before, $\dd\l_0=0$ and, hence, there is no zero-mode contribution to the second term. Thus\footnote{Note that \cite{FKZ} contains an equation that looks similar but is different since it is obtained by doing a first-order perturbation expansion in $m^2$ of the determinant.
}
\be\label{deltazeta2}
\dd\zeta(s)=\dd\wt\zeta(s)=2 s \int\d^2 x\sqrt{g}\  \dd\s(x) \big[ \wt\zeta(s,x,x)-m^2 \wt\zeta(s+1,x,x) \big] \ .
\ee
For $m\ne 0$, the term in brackets could have been  equally well written as $\zeta(s,x,x)-m^2 \zeta(s+1,x,x)$, but the writing in terms of the $\wt\zeta$ is valid for all non-zero and zero values of $m$. It follows that
\ba\label{deltazeta3}
\dd\zeta'(0)&=&2  \int\d^2 x\sqrt{g}\,  \dd\s(x)\, \wt\zeta(0,x,x)
-2 m^2 \int\d^2 x\sqrt{g}\,  \dd\s(x)\,  \lim_{s\to 0} \big[ \wt\zeta(s+1,x,x) + s \wt\zeta'(s+1,x,x)\big] 
\nonumber\\
\dd\zeta(0)&=&-2m^2 \int\d^2 x\sqrt{g}\  \dd\s(x) \lim_{s\to 0} \big[ s \wt\zeta(s+1,x,x)\big]
\ .
\ea
As recalled above, $\wt\zeta(s,x,x)$ has a pole at $s=1$ with residue $\frac{1}{4\pi}$, hence
\be\label{zetareg}
\wt\zeta(s,x,x)=\wt\zeta_{\rm reg}(s,x,x) +\frac{1}{4\pi(s-1)}
\quad\Rightarrow\quad
 \lim_{s\to 0} \big[ \wt\zeta(s+1,x,x) + s \wt\zeta'(s+1,x,x)\big] =  \wt\zeta_{\rm reg}(1,x,x) \ .
\ee
From \eqref{Gzetazetadef} one sees that $\wt\zeta_{\rm reg}(1,x,x)= \wt G_\zeta(x)-\frac{1}{4\pi} \ln \m^2$. Using also \eqref{zetaat0} we find\footnote{
The result for $\dd\zeta(0)$ also follows from \eqref{zetaat0} and the fact that $\int\d^2 x \sqrt{g} R=8\pi(1-h)$ is a topological invariant.
}
\ba\label{deltazeta4}
\dd\zeta'(0)&=&  \frac{1}{12\pi} \int\d^2 x\sqrt{g}\,  \dd\s(x)\, R(x)-\frac{\dd A}{A}
-2 m^2 \int\d^2 x\sqrt{g}\,  \dd\s(x)\, \big( \wt G_\zeta(x)+\frac{1}{4\pi} -\frac{1}{4\pi} \ln\m^2\big) \ , \nonumber\\
\dd\zeta(0)&=&-\frac{m^2}{2\pi} \int\d^2 x\sqrt{g}\  \dd\s(x)
\ .
\ea
Since $G_\zeta=\wt G_\zeta +\frac{1}{m^2 A}$, we arrive at two equivalent expressions for $\dd\zeta'(0)+\dd\zeta(0)\ln\m^2$~:
\ba\label{zetaprimezeta}
\dd\zeta'(0)+  \dd\zeta(0)\,\ln \m^2&=&\dd\wt\zeta'(0)+\dd\wt\zeta(0)\,\ln \m^2
\nonumber\\
&=&\frac{1}{12\pi} \int\d^2 x\sqrt{g}\,  \dd\s(x)\, R(x)-\frac{\dd A}{A}
-2 m^2 \int\d^2 x\sqrt{g}\,  \dd\s(x)\, \big( \wt G_\zeta(x)+\frac{1}{4\pi} \big) 
\nonumber\\
&=&\frac{1}{12\pi} \int\d^2 x\sqrt{g}\,  \dd\s(x)\, R(x)
-2 m^2 \int\d^2 x\sqrt{g}\,  \dd\s(x)\, \big( G_\zeta(x)+\frac{1}{4\pi} \big) 
\ .
\ea
As it stands, this result is exact in $m$ and holds whether $m^2 A$ is small or not. Let us insists that the $G_\zeta$ and $\wt G_\zeta$ appearing on the right-hand side are the massive ones. The first line is the appropriate way of writing to study the small $m^2$ asymptotics, as $\wt G_\zeta$ has a smooth limit for $m\to 0$.

\subsection{The massless case}

Let us quickly show, how in the massless case one recovers the well-known Liouville action. For $m=0$ one has $\dd\wt\zeta^{(0)}(0)=0$ and \eqref{zetaprimezeta} immediately gives
\be\label{zetaprimevarm0}
\dd\wt\zeta^{(0)'}(0)
= \frac{1}{12\pi} \int\d^2 x\sqrt{g}\,  \dd\s(x)\, R(x)-\frac{\dd A}{A}  \ .
\ee
Note that the last term precisely cancels the variation of the $-\frac{1}{2}\ln\frac{A}{A_0}$ in $S_{\rm grav}^{(0)}[g_0,g]$, cf \eqref{Sgravgen2}, and one gets the well-known result
\be\label{SgravLiouv}
\dd S_{\rm grav}^{(0)}[g_0,g]=-\frac{1}{24\pi} \int\d^2 x\sqrt{g}\,  \dd\s(x)\, R(x)
=-\frac{1}{24\pi} \dd S_L[g_0,\s] \  ,
\ee
cf.~eq.~\eqref{SMSAYvar}.
Of course, this is just the contribution of one conformal scalar field with $c=1$. The determinant that arises from fixing the diffeomorphism invariance  gives the same contribution but with a coefficient $+\frac{26}{24\pi}$, so that overall one gets 
\be\label{confmatplusghost}
S_{\rm grav}^{(0)}[g_0,g]\Big\vert_{\text{ghost + conf matter}}=\ \frac{26-c}{24\pi} \,S_L[g_0,\s]\ .
\ee

\subsection{The massive case}

Our starting point is obtained by combining eqs \eqref{Sgravgen2} and \eqref{zetaprimezeta}, using also \eqref{SgravLiouv}:
\be\label{Sgravvarm}
\dd S_{\rm grav}[g_0,g] 
=-\frac{1}{24\pi} \dd S_L[g_0,g]
+ m^2 \int\d^2 x\sqrt{g}\,  \dd\s(x)\, \big( G_\zeta(x)+\frac{1}{4\pi} \big) \ .
\ee
Our task then is to rewrite the second term on the r.h.s as the variation of some local functional.

Let us compute $\dd G_\zeta(x)$. In order to do so, we first establish a formula for $\dd G(x,y)$ under a variation $\dd g=2\dd\s\, g$ of the metric and thus under a corresponding variation $\dd\D_g=-2\dd\s\D_g$ of the Laplace operator. One can then either use the definition \eqref{Green} as an infinite sum and the perturbation theory formula \eqref{deltalambda}  and \eqref{deltaphi}, or directly use the defining differential equation \eqref{Green}. In any case one finds
\be\label{Gvar}
\dd G(x,y)=-2 m^2\int\d^2 z \sqrt{g}\, G(x,z)\, \dd\s(z)\, G(z,y) \ .
\ee
To obtain the variation of $G_\zeta$, according to \eqref{Gzetadef}  one needs to subtract the variation of the short-distance singularity. Now, the geodesic distance $\ell_g(x,y)$ transforms as (see e.g. appendix A1 of \cite{BF})
\be\label{geodistvar}
\dd \ell^2_g(x,y) =\ell_g^2(x,y) \big[ \dd\s(x) +\dd \s(y) +{\cal O}((x-y)^2)\big] \ .
\ee
It follows that
\be\label{geodistvar2}
\lim_{x\to y} \dd \ln \left[\m^2 \ell_g^2(x,y)\right]=2\,\dd\s(x) \ .
\ee
Plugging \eqref{Gvar} and \eqref{geodistvar2} into \eqref{Gzetadef} one gets
\be\label{Gzetavar}
\dd G_\zeta(x) =-2 m^2 \int\d^2 z \sqrt{g}\, \big(G(x,z)\big)^2\, \dd\s(z) + \frac{\dd\s(x)}{2\pi} \ .
\ee
Upon integrating this over $x$ one encounters 
\be\label{intermed1}
\int\d^2 x \sqrt{g} \big(G(x,z)\big)^2=\int\d^2 x \sqrt{g} \sum_{n,k\ge 0} \frac{\vf_k(x)\vf_k(z)\vf_n(x)\vf_n(z)}{\l_n\l_k}=\sum_{n\ge 0}\frac{\vf_n(z)\vf_n(z)}{\l_n^2}=\zeta(2,z,z) \ .
\ee
(Note that $\zeta(2,z,z)=G_2(z,z)$ is finite.) It follows that
\be\label{Gzetadeltasigma}
\dd\int \d^2 x\sqrt{g}\, G_\zeta(x) = 2 \int\d^2 x\sqrt{g}\,  G_\zeta(x) \dd\s(x)
- 2 m^2 \int\d^2 z \sqrt{g}\, \zeta(2,z,z) \dd\s(z) +\frac{1}{2\pi} \int \d^2 x \sqrt{g} \, \dd\s(x) \ .
\ee
One can then rewrite \eqref{Sgravvarm} as
\be\label{Sgravvarm1bis}
\dd S_{\rm grav}[g_0,g] 
= \dd  \left[-\frac{1}{24\pi}S_L[g_0,g]
+ \frac{m^2}{2}  \int\d^2 x\sqrt{g}\,  G_\zeta(x) \right]
+m^4 \int\d^2 x \sqrt{g} \, \zeta(2,x,x) \dd\s(x)
 \ .
\ee
Note  that in the second term, we can replace $G_\zeta$ by $\wt G_\zeta$ since the difference is 
$ \frac{m^2}{2}  \int\d^2 x\sqrt{g}\,  \frac{1}{m^2 A}=\frac{1}{2}$, whose variation vanishes.

Next, we use \eqref{zetaheat} to rewrite the last term as $m^4 \int_0^\infty \d t\, t \int\d^2 x \sqrt{g}\, K(t,x,x) \dd\s(x)$, and establish a formula for the variation of the integrated heat kernel $K(t)$. Since $\dd\l_0=0$ we have
\ba\label{heatvar}
\dd K(t)&=&\dd\wt K(t)=-\sum_{n> 0} t\, e^{-\l_n t} \dd\l_n
=2\sum_{n> 0} t \,e^{-\l_n t}(\l_n-m^2) \int\d^2 x\sqrt{g}\, \vf_n^2(x) \dd\s(x)
\nonumber\\
&=& -2 t \, \big(\frac{\d}{\d t} + m^2\big)\int\d^2 x\sqrt{g}\, \wt K(t,x,x)\dd\s(x)
=-2 t e^{-m^2 t}\, \frac{\d}{\d t} \int\d^2 x\sqrt{g}\, \wt K^{(0)}(t,x,x)\dd\s(x) \ ,
\nonumber\\
\ea
where we used \eqref{deltalambda} and \eqref{KmKo}. It then follows that
\ba\label{heatvar2}
&&\hskip-1.cm \frac{1}{2}\int_0^\infty \frac{\d t}{t} \big( e^{m^2 t} -m^2 t -1\big)  \dd \wt K(t)
=-\int_0^\infty  \d t  \big( e^{m^2 t} -m^2 t -1\big) 
\big(\frac{\d}{\d t} + m^2\big) \int\d^2 x\sqrt{g}\, \wt K(t,x,x)\dd\s(x) 
\nonumber\\
&=&m^4 \int_0^\infty \d t\, t  \int\d^2 x\sqrt{g}\, \wt K(t,x,x)\dd\s(x) 
=m^4 \int\d^2 x\sqrt{g}\, \wt\zeta(2,x,x)\dd\s(x) 
\nonumber\\
&=&m^4 \int\d^2 x\sqrt{g}\, \zeta(2,x,x)\dd\s(x) -\frac{1}{A} \int\d^2 x\sqrt{g}\, \dd\s(x) 
\ ,
\ea
where we integrated by parts and used \eqref{zetaheat}. The boundary terms do not contribute\footnote{
Had we started with $\dd K$ rather than $\dd \wt K$ and written this equation for $K$ and $\zeta(2,x,x)$, the $-\frac{1}{A}\int\sqrt{g}\dd\s$ would have appeared as the boundary term.} 
since $\wt K$ vanishes at $t=\infty$ as $e^{-\l_1 t}$ and $\l_1-m^2>0$.  Upon inserting this into \eqref{Sgravvarm1bis}
we finally get
\ba\label{Sgravvarm3}
\dd S_{\rm grav}[g_0,g] 
= \dd  \Bigg[-\frac{1}{24\pi}S_L[g_0,g]+\frac{1}{2}\ln\frac{A}{A_0}
+ \frac{m^2}{2}  \int\d^2 x\sqrt{g}\,  \wt G_\zeta(x) 
\nonumber\\
\ \ +\ \frac{1}{2}\int_0^\infty \frac{\d t}{t} \big( e^{m^2 t} -m^2 t -1\big)  \wt K(t) \Bigg]\, .
\ea
Note that in the last term the $t$-integral is convergent both at $t=0$ and at $t=\infty$. This is immediately integrated as
\ba\label{Sgravvarm3bis}
S_{\rm grav}[g_0,g] 
=-\frac{1}{24\pi}S_L[g_0,g]+\frac{1}{2}\ln\frac{A}{A_0}
+ \frac{m^2}{2}  \int\d^2 x\big( \sqrt{g}\, \wt G_\zeta(x;g)- \sqrt{g_0}\, \wt G_\zeta(x;g_0)\big) 
\nonumber\\
\ \ +\ \frac{1}{2}\int_0^\infty \frac{\d t}{t} \big( e^{m^2 t} -m^2 t -1\big) \big( \wt K(t;g) -\wt K(t;g_0)\big) \, .
\ea

Thus we have expressed the gravitational action $S_{\rm grav}[g_0,g]$ in terms of (local) functionals of $g$ and $g_0$ that are all perfectly well-defined without any need of analytical continuation (contrary to the initial $\zeta'(0)$). Recall also from our  remark at the end of sect.~\ref{massivemassless} that, if we define $S_{\rm grav}[g_0,g]$ through the spectral cutoff regularization of the logarithm of the determinant, the r.h.s. also involves the variation of the cosmological constant action
$\m_0 A$ with a coefficient $\m_0\sim \L^2 ( \ln \frac{\l^2}{\m^2} + {\rm const})$, to be cancelled by a corresponding counterterm.

With view on the small mass expansion studied below, it  will be useful to rewrite \eqref{Gzetavar} to obtain the variation of $\wt G_\zeta$ in terms of quantities that all have well-defined limits as $m\to 0$.
Recall that $G=\frac{1}{m^2 A}+\wt G$ and $G_\zeta=\frac{1}{m^2 A}+\wt G_\zeta$. Thus \eqref{Gzetavar} can be rewritten as
\be\label{Gzetatildevar}
\dd \wt G_\zeta(x) =\frac{\dd\s(x)}{2\pi} -\frac{4}{A} \int\d^2 z\sqrt{g}\, \wt G(x,z)\dd\s(z)
-2 m^2 \int\d^2 z \sqrt{g}\, \big(\wt G(x,z)\big)^2\, \dd\s(z)  \ .
\ee
Using \eqref{deltasigmadeltaphi}, we can express $\dd\s$ in the second term as $-A\D\dd\f/4$ (the $\dd A/2A$ piece doesn't contribute since $\wt G$ has no zero-mode). Using the differential equation \eqref{GminusGtilde} satisfied by $\wt G$, one finds for the second term
\ba\label{intermediateeq}
-\frac{4}{A} \int\d^2 z\sqrt{g}\, \wt G(x,z)\dd\s(z)
=\int\d^2 z\sqrt{g}\, \dd\f(z)\, \D_z\wt G(x,z)
\nonumber\\
=\dd\f(x) -\dd S_{\rm AY}[g_0,g]- m^2 \int\d^2 z\sqrt{g}\, \wt G(x,z) \dd\f(z) \ ,
\ea
with $\dd S_{\rm AY}$ given in \eqref{SMSAYvar}. Thus
\be\label{Gzetatildevar2}
\dd \wt G_\zeta(x) =\frac{\dd\s(x)}{2\pi} 
+\dd\f(x) -\dd S_{\rm AY}[g_0,g]
- m^2 \int\d^2 z \sqrt{g}\, \left[ \wt G(x,z) \dd\f(z)  + 2 \big(\wt G(x,z)\big)^2\, \dd\s(z)\right]  \ .
\ee
In exactly the same way we also get
\ba\label{Gtildevar2}
\dd \wt G(x,y) &=&
\frac{1}{2}\dd\f(x)+  \frac{1}{2}\dd\f(y) -\dd S_{\rm AY}[g_0,g]
\nonumber\\
&&- m^2 \int\d^2 z \sqrt{g}\, \left[ \frac{1}{2}\wt G(x,z) \dd\f(z)  +  \frac{1}{2}\wt G(y,z) \dd\f(z) + 2 \wt G(x,z)\, \dd\s(z) \wt G(z,y)\right]  \ .
\ea
While these are  exact relations valid for all $m$, they are written in a way that makes the small mass expansions obvious.
 
\subsection{Small mass expansion}

Equations \eqref{Sgravvarm3} and \eqref{Sgravvarm3bis} are non-perturbative in $m$. However, they are also written in a way that immediately allows for a perturbative expansion in $m$, since $\wt G_\zeta$ and $\wt K$ have smooth limits as $m\to 0$. The $m^2\to 0$ limits of \eqref{Sgravvarm3} and \eqref{Sgravvarm3bis} also exhibit an extra term  $\frac{1}{2} \ln \frac{A}{A_0}$  which only gets removed for $m^2=0$ due to the difference in the definitions \eqref{Sgravgen2}.

\vskip3.mm
\noindent{\bf Order $m^2 A$ contributions}
 
\vskip3.mm
\noindent
Let us first give the order $m^2 A$-correction $S_{\rm grav}^{(1)}$ to $S_{\rm grav}^{(0)}$. Since $\wt K(t)=e^{-m^2 t} \wt K^{(0)}(t)$ and $\l_n^{(0)}\sim \frac{1}{A}$ it follows that 
\be\label{Kpiecem4}
\int_0^\infty \frac{\d t}{t} \big( e^{m^2 t} -m^2 t -1\big)  \wt K(t)={\cal O}((m^2 A)^2)\ .
\ee
Next, it follows from \eqref{GminusGtilde} and \eqref{massiveGexp}, upon letting $x\to y$ and subtracting the short-distance singularity, that
\be\label{Gzetatildeexp}
\wt G_\zeta(x)=\wt G_\zeta^{(0)}(x) +\sum_{r=1}^\infty (-m^2)^r \wt G_{r+1}^{(0)}(x,x)
\ ,
\ee
so that $\wt G_\zeta(x)=\wt G_\zeta^{(0)}(x)+{\cal O}(m^2 A)$.
Thus the term of order $m^2 A$ of the gravitational action can be read from \eqref{Sgravvarm3bis} as
\be\label{Sgravvar1}
S_{\rm grav}^{(1)}[g_0,g] =    \frac{m^2}{2} \Big( A \Psi_G[g]-A_0 \Psi_G[g_0] \Big)
= \frac{m^2 A}{2} \Big( \Psi_G[g]-\Psi_G[g_0]\Big) + (A-A_0) \frac{m^2}{2}\Psi_G[g_0]\ ,
\ee
where, following \cite{FKZ}, we have introduced
\be\label{PsiG}
\Psi_G[g]=\frac{1}{A}\int\d^2 x\sqrt{g}\,  \wt G_\zeta^{(0)}(x;g) \ .
\ee
The variation of $\wt G_\zeta$ was given in \eqref{Gzetatildevar2} and that of $\wt G_\zeta^{(0)}$ immediately follow as
\be\label{Gzetatildem0var}
\dd \wt G_\zeta^{(0)}(x) =\frac{\dd\s(x)}{2\pi} 
+\dd\f(x) -\dd S_{\rm AY}[g_0,g] \ ,
\ee
so that
\be\label{Gzetam0gg0}
\wt G_\zeta^{(0)}(x;g) =\wt G_\zeta^{(0)}(x;g_0)+ \frac{\s(x)}{2\pi} 
+\f(x) -S_{\rm AY}[g_0,g]  \ .
\ee
This relation has been derived before in \cite{FKZ}. Using \eqref{gg0sig}, it it then straightforward to obtain (\cite{FKZ})
\be\label{intGzetavar}
\Psi_G[g]-\Psi_G[g_0]
=\frac{1}{8\pi} \int \d^2 x\sqrt{g_0}\, \Big[ \frac{4}{A} \s e^{2\s} -2\pi \f \D_0\f -4\pi \f \, \D_0 \wt G_\zeta^{(0)}(x;g_0)\Big] \ .
\ee
For genus $h=0$ and choosing  $g_0$ to be the round metric on the sphere, $\wt G_\zeta^{(0)}(x;g_0)$ is a constant, and one directly gets the Mabuchi action for $h=0$.
More generally, for arbitrary genus, \cite{FKZ} show that
\be\label{intGzetavargenh}
\Psi_G[g]-\Psi_G[g_0]
=\frac{1}{8\pi} S_{\rm M}[g_0,g] +  h \Big( S_{\rm AY}[g_0,g] - \int \d^2 x\sqrt{g_c}\, \f\Big) \, ,
\ee
where $g_c$ is the canonical metric on the Riemann surface. Finally, \eqref{Sgravvar1} becomes
\be\label{Sgravvar2}
S_{\rm grav}^{(1)}[g_0,g] =    \frac{m^2 A}{2} \Big[ \frac{1}{8\pi} S_{\rm M}[g_0,g] +  h \Big( S_{\rm AY}[g_0,g] - \int \d^2 x\sqrt{g_c}\, \f \Big)   \Big]
+(A-A_0) \frac{m^2}{2}  \Psi_G[g_0]\ .
\ee
While the last term  contributes to the cosmological constant action,
the other terms are to be considered as the genuine order $m^2 A$ correction to the gravitational action, and it involves the Mabuchi and Aubin-Yau actions.

\vskip3.mm
\noindent{\bf Higher-order contributions $S_{\rm grav}^{(r)}$ for $r\ge 2$}
 
\vskip3.mm
\noindent
It is straighforward to obtain the expansion in powers of $m^2$ of the terms in \eqref{Sgravvarm3} or \eqref{Sgravvarm3bis}. Denoting the term $\sim m^{2r}$ by $S_{\rm grav}^{(r)}$, we obviously have
\be\label{Sgravvarm4}
S_{\rm grav}[g_0,g] 
=\sum_{r=0}^\infty \, S_{\rm grav}^{(r)}[g_0,g] \ .
\ee
In particular, using $\wt K(t)=e^{-m^2 t} \wt K^{(0)}(t)$, one has
\ba\label{Ktermexp}
&&\hskip-1.cm \frac{1}{2}\int_0^\infty \frac{\d t}{t} \big( e^{m^2 t} -m^2 t -1\big)  \wt K(t) 
=\sum_{r=2}^\infty (-m^2)^r \frac{(r-1)}{2 r!} \int_0^\infty \d t\, t^{r-1} \wt K^{(0)}(t)
\nonumber\\
&&=\sum_{r=2}^\infty (-m^2)^r \frac{(r-1)}{2 r} \wt\zeta^{(0)} (r)
=\sum_{r=2}^\infty (-m^2)^r \frac{(r-1)}{2 r} \int\d^2 x \sqrt{g}\, \wt G^{(0)}_r(x,x) \ ,
\ea
and combining this with \eqref{Gzetatildeexp}, eq.~\eqref{Sgravvarm3bis} yields
\be\label{Sgravvar-r}
S_{\rm grav}^{(r)}[g_0,g] = \frac{(-)^{r+1}}{2r} m^{2r}\, \big(\wt\zeta^{(0)}(r;g)-\wt\zeta^{(0)}(r;g_0)\big)
\quad , \quad r\ge 2 \ .
\ee
Of course, this coincides with the result of the naive expansion of $\sum_{n\ge 1} \ln \l_n =\sum_{n\ge1} \ln (\l_n^{(0)} + m^2)$  in powers of $m^2$. One may  rewrite $S_{\rm grav}^{(r)}[g_0,g] $ as a local functional:
\be\label{Sgravvar-r2}
S_{\rm grav}^{(r)}[g_0,g] = \frac{(-)^{r+1}}{2r} m^{2r}\, \Big[\int\d^2 x \sqrt{g}\, \wt G_r^{(0)}(x,x;g)
-\int\d^2 x \sqrt{g_0}\, \wt G_r^{(0)}(x,x;g_0) \Big]
\quad , \quad r\ge 2 \ .
\ee

While these $S_{\rm grav}^{(r)}[g_0,g] $ are appropriate local gravitational actions, it would be desirable to express them in terms of more geometric quantities like the conformal factor or the K\"ahler potential, as was the case for $S_{\rm grav}^{(0)}[g_0,g]$ and $S_{\rm grav}^{(1)}[g_0,g] $ with the Liouville, Mabuchi and Aubin-Yau actions. To our knowledge, there does not seem to exist any appropriate functional in the mathematical literature. Nevertheless, since the $\wt G_r^{(0)}$ are entirely determined in terms of the properties of the Riemann surface, they are purely geometric quantities. 

\vskip5.mm
\noindent
{\Large\bf Acknowledgements}

\vskip3.mm
\noindent
L.L. is supported by the German Excellence Initiative at the University of Cologne. During the initial phase of this work at the ENS, L.L. was  supported by a fellowship from the Capital Fund Management foundation and  the Foundation of the Physics Department DEPHY.
\vskip5.mm



\begin{thebibliography}{99}

\bibitem{Liouville1}
{A.M.~Polyakov,
\emph{Quantum geometry of bosonic strings},
Phys.\ Lett. {\bf B103}, 207 (1981).}
 
 \bibitem{Mabuchi}
 {T.~Mabuchi, {\it K-energy maps integrating Futaki invariants,} T\^ohuku\ Math.\ J.\ {\bf 38} (1986) 575--593.\\
T.~Mabuchi, {\it Some symplectic geometry on compact K\"ahler manifolds,} Osaka J. Math. {\bf 24} (1987) 227--252;\\ S.~Semmes, {\it Complex Monge-Amp\`ere and symplectic manifolds,} Amer.\ J.\ Math.\ {\bf 114} no. 3 (1992) 495--550.}

\bibitem{AubinYau} 
{D.H.~Phong, J.~Sturm, {\it Lectures on stability and constant scalar curvature},  arXiv:0801.4179 [math.DG] (2008).}

\bibitem{FKZ}
{F.~Ferrari, S.~Klevtsov and S.~Zelditch,
\emph{Gravitational actions in two dimensions and the Mabuchi functional}, 
Nucl.\ Phys. {\bf B859}, 341 (2012), arXiv:1112.1352 [hep-th].}

\bibitem{BFK}
{A. Bilal, F. Ferrari and S. Klevtsov,
\emph{2D quantum gravity at one loop with Liouville and Mabuchi actions}, Nucl. Phys. {\bf B880} (2014) 203, arXiv:1310.1951 [hep-th].}

\bibitem{BL}
{A. Bilal and L. Leduc, \emph{Liouville and Mabuchi quantum gravity at two and three loops}, to appear.}

\bibitem{BF}
{A. Bilal and F. Ferrari, \emph{Multi-loop zeta function regularization and spectral cutoff in curved spacetime}, Nucl. Phys. {\bf B877} (2013) 956, arXiv:1307.1689 [hep-th].}

\end{thebibliography}
\end{document}